\documentclass[aps,prl,twocolumn,showpacs,amsmath,amssymb,floatfix,superscriptaddress]{revtex4-1}

\usepackage{graphicx}
\usepackage{hyperref}
\usepackage{color}
\usepackage[yyyymmdd,hhmmss]{datetime}
\usepackage[normalem]{ulem}
\usepackage{multirow}
\usepackage{float}
\usepackage{array, blindtext}
\usepackage{adjustbox}
\usepackage{makecell}
\usepackage{bm}

\begin{document}
\title{In-situ Control of the Photon-Magnon Synchronized Mode in a Miniaturized Magnetometer}

\author{Yuting Shi}
\affiliation{School of Physics and State Key Laboratory of Crystal Materials, Shandong University, No. 27 Shandanan Road, Jinan, 250100 China}

\author{Xudong Wang}
\affiliation{School of Physics and State Key Laboratory of Crystal Materials, Shandong University, No. 27 Shandanan Road, Jinan, 250100 China}

\author{Yue Zhao}
\affiliation{School of Physics and State Key Laboratory of Crystal Materials, Shandong University, No. 27 Shandanan Road, Jinan, 250100 China}

\author{Lijun Yan}
\affiliation{School of Physics and State Key Laboratory of Crystal Materials, Shandong University, No. 27 Shandanan Road, Jinan, 250100 China}

\author{Bimu~Yao}
\affiliation{State Key Laboratory of Infrared Physics, Shanghai Institute of Technical Physics, Chinese Academy of Sciences, Shanghai 200083, China}

\author{Shishen Yan}
\affiliation{School of Physics and State Key Laboratory of Crystal Materials, Shandong University, No. 27 Shandanan Road, Jinan, 250100 China}

\author{Lihui Bai} \email{lhbai@sdu.edu.cn;}
\affiliation{School of Physics and State Key Laboratory of Crystal Materials, Shandong University, No. 27 Shandanan Road, Jinan, 250100 China}

\author{Jinwei Rao} \email{raojw@sdu.edu.cn} 
\affiliation{School of Physics and State Key Laboratory of Crystal Materials, Shandong University, No. 27 Shandanan Road, Jinan, 250100 China}

\begin{abstract}
Photon–magnon synchronized modes (PMSMs) provide a route for magnetic-field sensing by exploiting their enhanced field responsivity due to cusp points (CPs). However, the previous PMSM magnetometer relies on bulky discrete components and lacks in situ control of photon–magnon interactions, limiting its integration and adaptability. Here, we demonstrate a miniaturized PMSM magnetometer integrated on a printed circuit board. A variable resistor is used to tune the photon–magnon coupling strength in situ. By adjusting the variable resistor, the electromagnetic boundary condition of the magnetometer is altered; hence, the reservoir-mediated photon-magnon interaction is changed. The coupling strength shows a non-monotonic dependence on the reflection coefficient adjusted by the variable resistor and reaches its minimum at the impedance matching condition. Furthermore, we develop a non-Hermitian model with standing-wave interference to describe the PMSM dispersion and derive the critical condition for CP formation. This work realizes a compact and tunable PMSM magnetometer and provides a pathway for controlling photon–magnon interactions in integrated magnetic sensing systems.
\end{abstract}

\maketitle
Magnetic field sensing is a fundamental technology with a broad range of applications\cite{RevModPhys.89.035002, 1634415,5443656, Khan_2021}, spanning from geophysical exploration and space navigation\cite{connerney2017the, 10.1190/1.2133784, s24217047} to biomedical imaging\cite{aslam2023quantum, RevModPhys.65.413, BROOKES2022621} and non-destructive testing\cite{pmid36295427, WANG2012382, INESSILVA2023101155}. In recent decades, the demand for high-precision magnetometry has driven the development of various sensing platforms, including superconducting quantum interference devices\cite{granata2016nano, Vasyukov2013, deGraaf2018}, atomic magnetometers\cite{PhysRevLett.89.130801, budker2007optical}, and diamond-based nitrogen-vacancy center magnetometers\cite{RevModPhys.92.015004, PhysRevResearch.2.023394, RevModPhys.92.015004, Casola2018}. To extend the boundaries of sensing performance, researchers have increasingly focused on hybrid architectures to achieve enhanced capabilities that overcome the detection limits of individual subsystems. Among these, microwave-based magnonic sensors\cite{doi:10.1126/science.aaa3693,Wang2024, PhysRevLett.132.116701, PhysRevLett.120.057202, PhysRevLett.123.107701, PhysRevLett.130.046705, YUAN20221} have emerged as a promising frontier due to their high sensitivity, room-temperature operation, and ease of integration\cite{dkwj-jv6l, 10.1063/5.0024369}.

As a representative example, in a recent study, the cusp point (CP) of the photon-magnon synchronized mode (PMSM) was realized in an active cavity magnonic system\cite{dkwj-jv6l}. It exhibits an ultra-steep nonlinear response to external magnetic field variations, thereby enabling enhanced detection sensitivity for weak magnetic signals. This mechanism combines the high frequency resolution of active photon modes with the intrinsic magnetic responsivity of magnon modes, effectively overcoming the limitations of each individual subsystem\cite{PhysRevLett.131.106702, dkwj-jv6l}. Nevertheless, the existing PMSM magnetometer is still primarily based on bulky assemblies of discrete components, lacking compact integration and in situ tunability of the coupling strength. Such limitations restrict adaptive control of the synchronization condition and hinder practical deployment in scalable sensing applications.

\begin{figure}[htbp]
\includegraphics[width = 8.17 cm, trim=0.78cm 0.5cm 0.38cm 0.8cm, clip]{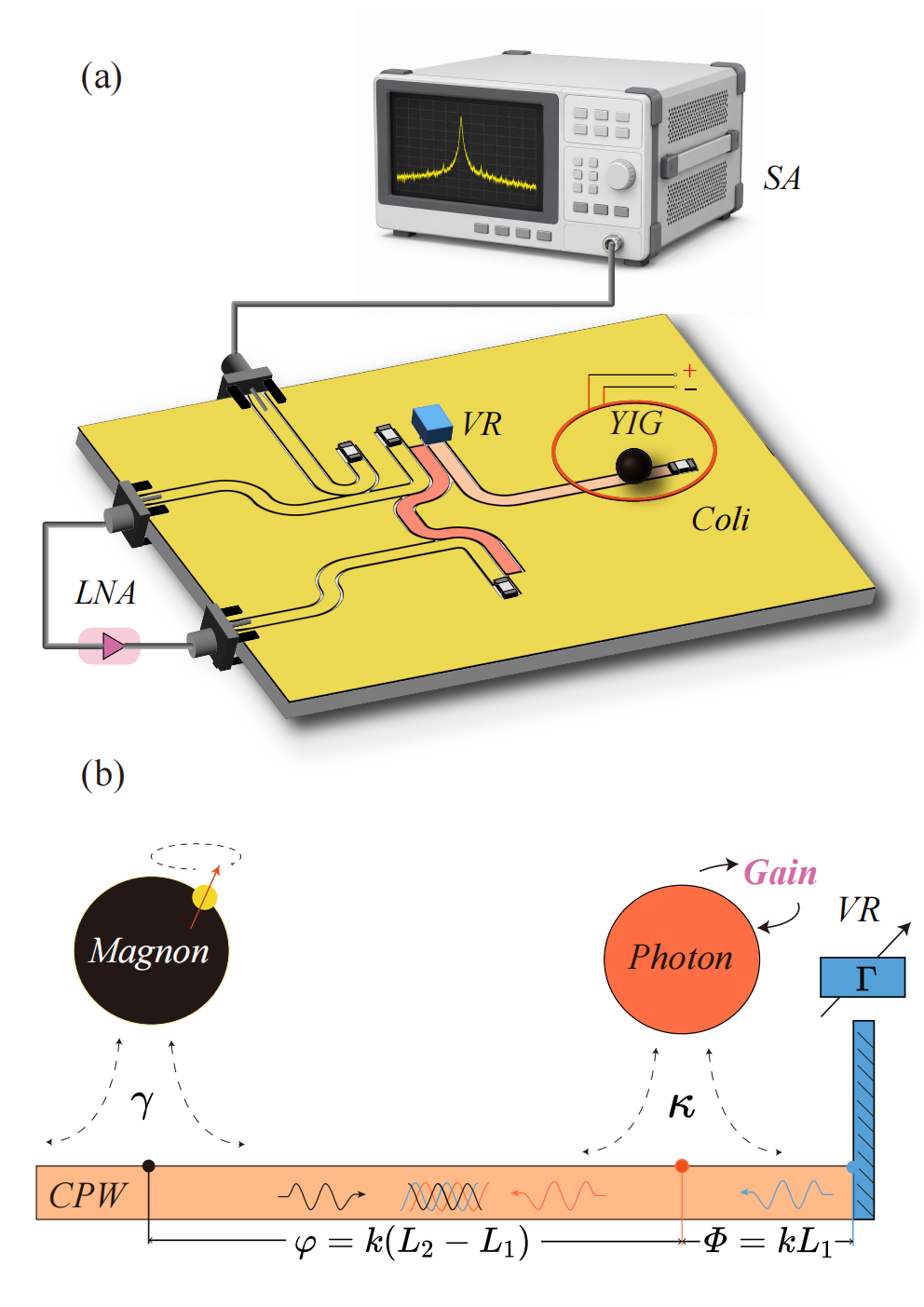}
\caption{
(a) Schematic of the integrated planar magnetometer. The blue component highlights the VR positioned at the waveguide end, which is utilized for in-situ tuning of the terminal impedance (reflection coefficient $\Gamma$) to modulate the standing-wave environment. A low-noise amplifier (LNA) compensates for cavity damping, while a YIG sphere interacts with the CPW, and a SA monitors the output signal.(b) Schematic of the coupling effect in the PMSM system. $\varPhi$ represents the microwave propagation phase from the cavity to the VR, and $\varphi$ denotes the relative phase shift between the photon and magnon modes
}
\label{fig1}
\end{figure}

To address these challenges, in this work, we design a miniaturized PMSM magnetometer based on a compact integrated printed circuit board (PCB) architecture. By adopting this platform, we reduce the system's footprint while enabling in situ tuning of the PMSM via a variable resistor (VR), providing a robust solution for portable magnetic sensing. In addition to the experimental setup, we develop a theoretical model to elucidate the underlying mechanism and derive the critical condition required to reach the CP singularity. Our work demonstrates a viable approach to miniaturizing the PMSM-based magnetometer and advances the study of CP singularities in nonlinear cavity magnonic systems.

Our PMSM magnetometer is fabricated on a Rogers 4350B PCB with dimensions of $45.5 \times 37.0 \times 0.76 \text{ mm}^3$ (Fig. \ref{fig1} (a)). The PCB comprises two functional sections. The left section integrates a passive planar cavity and three ports: two ports are connected to an external amplifier to form a feedback gain loop that enables self-oscillation of the cavity photon mode, while the third port is linked to a signal analyzer (SA) to detect the emission spectrum of the PMSM. The right section features a coplanar waveguide (CPW) coupled to the cavity. The end of the CPW near the cavity is integrated with a VR, whereas its far end is soldered to a $50\,\Omega$ resistor for impedance matching. The VR adjusts the terminal resistance $R$ of the CPW, enabling in situ control of the impedance conditions of microwaves. A yttrium iron garnet (YIG) sphere with a diameter of $1.2$ mm is mounted on the CPW at a precisely designed separation from the cavity. Beneath the PCB, a permanent magnet provides a stable static bias magnetic field, while a surrounding solenoid coil allows for precise fine-tuning.

As we adjust the resistance $R$ of the VR, the resulting impedance mismatch creates a reflective boundary and hence alters the electromagnetic environment of both the cavity and the YIG sphere. The photon-magnon coupling status in our device evolves from a purely traveling-wave-mediated regime into a hybrid regime that incorporates both traveling and standing waves. This transition modifies the indirect coupling strength and also shifts the relative phase between the photon mode and the magnon mode. To compensate for this phase deviation in the experiment, we finely adjust the position of the YIG sphere to offset the phase shift induced by the reflective boundary. With this compensation, we can consider the coupling strength ($|g|/2\pi$) as a function of the reflection coefficient $\Gamma$ ($\Gamma = (R - Z_0)/(R + Z_0)$)\cite{pozar2011microwave} at the VR. Here, $Z_0$ represents the characteristic impedance of the CPW, which is designed to be $50\ \Omega$ in this device.

To theoretically describe our device, we construct an effective Hamiltonian model that has the following form\cite{Haroche2006Exploring}:
\begin{equation}
\begin{aligned}
\mathcal{H}/\hbar &= \mathcal{H}_0/\hbar-i\Lambda
= \begin{bmatrix}\omega_{p}-i(\beta-G)& 0\\ 0&\omega_{m}-i\alpha\end{bmatrix}\\&-i\begin{bmatrix}\kappa(1 + \Gamma e^{i2\varPhi})& \sqrt{\kappa \gamma} e^{i \varphi} (1 + \Gamma e^{i 2\varPhi})\\ \sqrt{\kappa \gamma} e^{i \varphi} (1 + \Gamma e^{i 2\varPhi})&\gamma(1 - \Gamma e^{i2(\varPhi+\varphi)})\end{bmatrix},
\end{aligned} \label{equation1}
\end{equation}
where $\omega_p$ and $\omega_m$ denote the photon and magnon mode frequencies, respectively. $\kappa$ and $\gamma$ are the radiation rates of the photon and magnon modes into the CPW, while $\beta$ and $\alpha$ correspond to the intrinsic damping rates of the photon and magnon modes. The nonlinear gain acting on the cavity photon mode is given as $G=N(1-i\sqrt{\epsilon/2}\hat{a}-\epsilon \left|\hat{a} \right|^2)$, where $\hat{a}$ is the annihilation operator of the cavity photon mode, and $N$ and $\epsilon$ represent the linear negative damping and the saturation factor, respectively. $\varPhi$ represents the microwave propagation phase from the cavity to the VR, while $\varphi$ denotes the relative phase shift between the photon and magnon modes.

\begin{figure}[htbp]
\includegraphics[width = 8.17 cm, trim=0.75cm 0.5cm 0.65cm 0.6cm, clip]{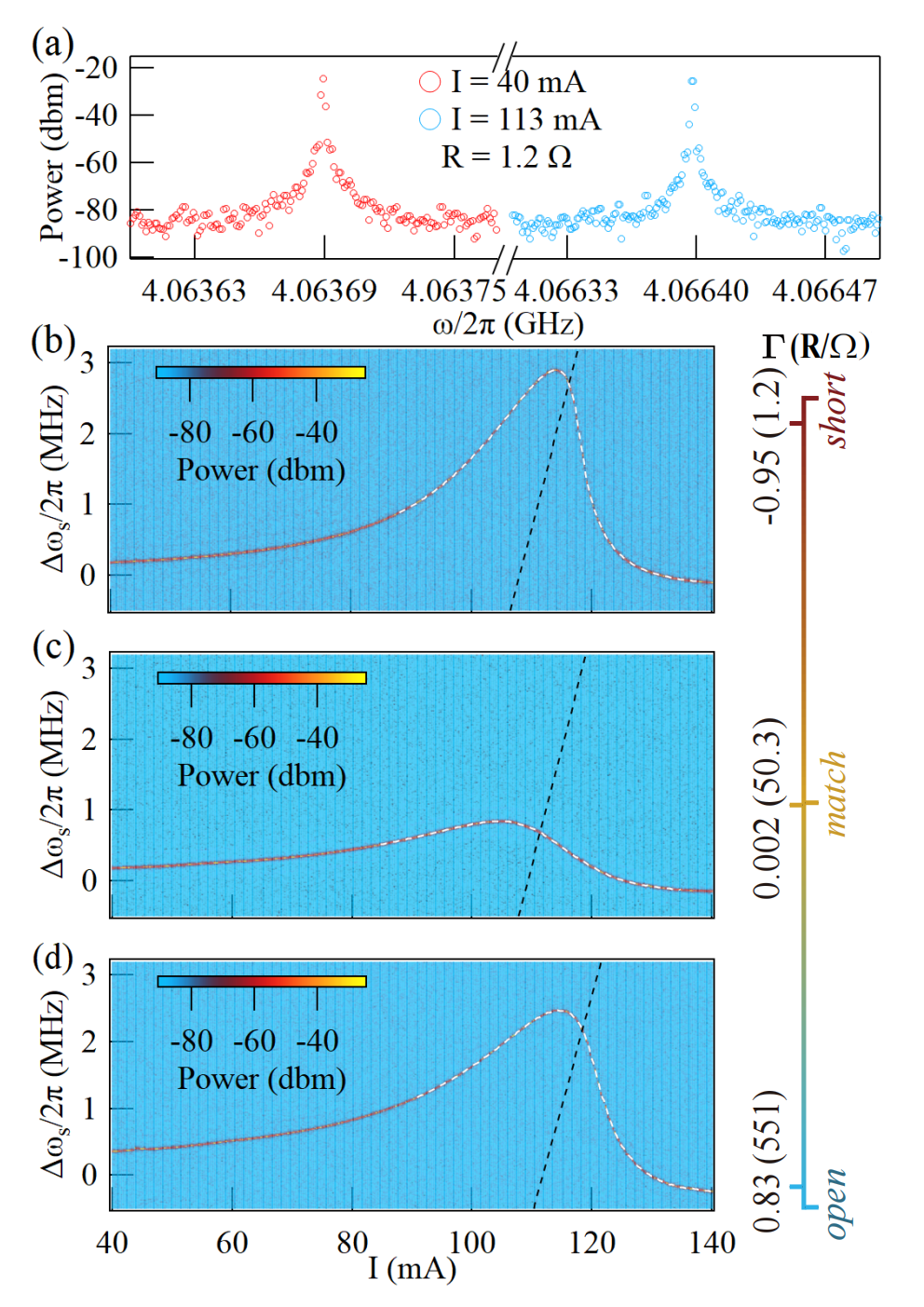}
\caption{
(a) Radiation spectra of the system measured by the SA under two different coil currents at a fixed terminal resistance. (b-d) Mappings of the magnetometer as a function of the coil current, measured under different terminal resistances (from top to bottom: 1.2 $\Omega$, 50.3 $\Omega$, and 551 $\Omega$, respectively) while maintaining the system at the same. The white dashed curves represent the theoretical fitting results based on Eq. \ref{equation3}.
}
\label{fig2}
\end{figure}

The first term of Eq. (\ref{equation1}), $\mathcal{H}_0$, corresponds to the uncoupled two modes, and the second term ($\Lambda$) describes the reservoir-mediated effect. The diagonal elements of $\Lambda$ represent the reservoir‑induced self‑energy shifts. They describe the process in which microwave radiations emitted by the cavity and magnon mode are reflected at the VR and then reabsorbed by themselves. The cavity photon mode couples predominantly to the electric field of the CPW, while the magnon mode couples to the magnetic field. Dictated by electromagnetic boundary conditions, the corresponding magnetic reflection coefficient is intrinsically $-\Gamma$, thereby introducing a critical $\pi$ phase shift into the magnon self-energy. By incorporating these external reservoir contributions, the total magnon damping rate becomes $\alpha_{t} = \alpha + \gamma(1 - \Gamma \cos [2(\varPhi+\varphi)])$.

The off-diagonal elements $\Lambda_{12}=\Lambda_{21}=\sqrt{\kappa \gamma} e^{i \varphi} (1 + \Gamma e^{i 2\varPhi})$ describe the reservoir-mediated coupling effect, representing the energy exchange rate between the photon mode and the magnon mode. The relationship between the coupling strength $|g|$ and $\Gamma$ is derived as:
\begin{equation}\label{equation4}
|g|=\sqrt{\kappa \gamma (1+|\Gamma|^2+2|\Gamma|\cos{2\varPhi})}.
\end{equation}
In the limiting case of a pure traveling-wave assumption with $\Gamma = 0$, the interaction term effectively reduces to the simplified form $i\sqrt{\kappa\gamma}e^{i\varphi}$. 

When the gain surpasses the damping rate of the cavity photon mode, the system enters a synchronized state where the photon and magnon modes oscillate at an identical frequency $\omega_{s}$ with a constant phase difference $\theta$, i. e., $\hat{a}=\left |\hat{a} \right |e^{-i\omega_{s}t}$ and $\hat{m}=\left |\hat{m} \right |e^{-i\omega_{s}t-\theta}$\cite{PhysRevLett.131.106702}. From the dynamically coupled equations, we can obtain a cubic equation for the PMSM detuning defined as $\Delta\omega_s = \omega_s - \omega_p$:
\begin{equation}\label{equation3}
\Delta\omega_s^3 + a_2 \Delta\omega_s^2 + a_1 \Delta\omega_s + a_0 = 0.
\end{equation}
The polynomial coefficients are explicitly expressed as :
\begin{equation}
\begin{aligned}
 a_0& = -|g|^2\cos(2\varphi')(\omega_m - \omega_p) - |g|^2\alpha_t\sin(2\varphi'),\\
 a_1& = (\omega_m - \omega_p)^2 + \alpha_t^2 + |g|^2\cos(2\varphi'),\\
 a_2& = -2(\omega_m - \omega_p).
\end{aligned}  
\end{equation}
Here, $\varphi'=\varphi+\arctan[\Gamma\sin{\varPhi}/(1+\Gamma\cos{\varPhi})]$ is an effective phase. Adjusting the reflection coefficient $\Gamma$ can change $\varphi'$ and hence affect the values of $a_1$ and $a_0$. In contrast, $a_2$ is twice the magnon detuning (i.e., $\Delta_m=\omega_m-\omega_p$). It is independent of $\Gamma$ and can be tuned by the magnetic field via $\omega_m$. Therefore, either the magnetic field or the reflection coefficient can be used to adjust these cubic coefficients, providing cooperative control over the PMSM. In our magnetometer, the tunability of the magnetic field is leveraged to sense magnetic field variations, while the tunability of the reflection coefficient is used to tune the PMSM toward optimal sensing performance.

\begin{figure}[htbp]
\centering
\includegraphics[width = 8.1 cm, trim=0.50cm 0.5cm 0.8cm 0.75cm, clip]{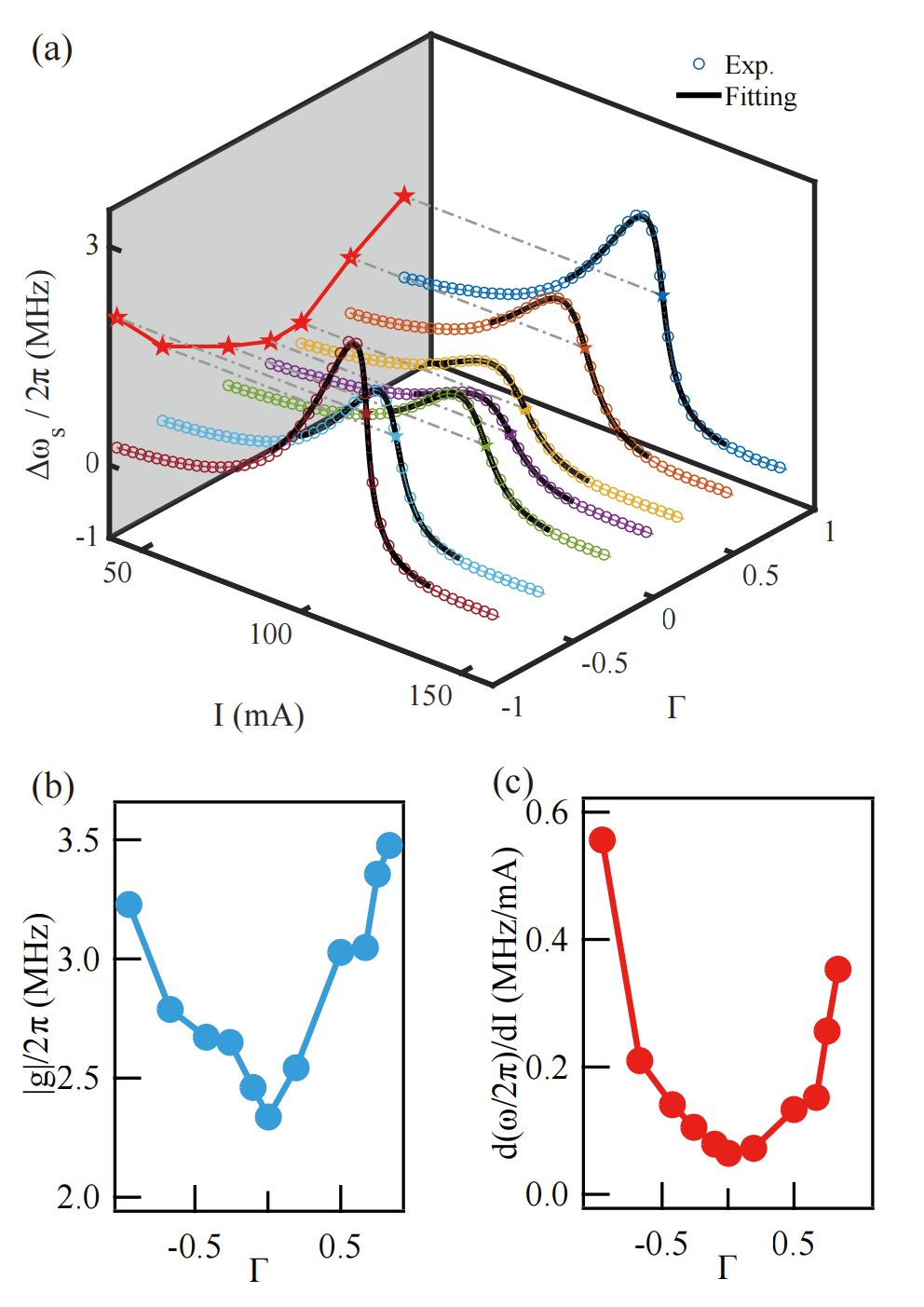}
\caption{
(a) Extracted frequency detuning $\Delta\omega_{s}/2\pi$ as a function of the coil current $I$ for different reflection coefficients $\Gamma$. Hollow circles represent the experimental data, and solid black lines denote the corresponding fits using Eq.~\ref{equation3}. The detuning values corresponding to the extrema of the first derivative of each dispersion curve are highlighted (red stars connected by a dashed line) and projected onto the $YOZ$ plane to illustrate the evolution. 
(b) Extracted effective coupling strength $|g|/2\pi$ as a function of $\Gamma$. 
(c) Maximum slope of the frequency shift with respect to the coil current, $d(\omega/2\pi)/dI$, as a function of $\Gamma$.
}
\label{fig3}
\end{figure}

Figure \ref{fig2} (a) presents a comparison of two emission spectra of our device measured at different magnetic fields. As the coil current is tuned, the magnetic field experienced by the YIG sphere changes. Consequently, the resonant peak in the emission spectrum shifts to a higher frequency, demonstrating the magnetic-field-dependent response of the PMSM. To further characterize the in situ tunability of our device with respect to the magnetic field and the reflection coefficient, we performed measurements at three different VR resistance values. Figures \ref{fig2}(b)–(d) show the corresponding spectral mappings versus coil current for these three resistance values. All measurements are performed under the condition of an identical effective phase $\varphi'$, ensuring a fixed interference configuration. In Fig. \ref{fig2}(b), the VR is set at $R = 1.2\ \Omega$, corresponding to shorting the terminal of the CPW with a reflection coefficient of $\Gamma=-0.95$. The red color indicates the emission peak of the PMSM, while the black dashed line represents the magnon detuning ($\Delta_m$) at different coil currents. As $\Delta_m$ increases, the PMSM peak first shifts to its highest frequency and then drops rapidly in frequency. Using our theoretical model, we can extract the photon-magnon coupling strength from the PMSM detuning variation with the coil current. The extracted value is $g/2\pi=3.27$ MHz. When the VR is set at $R = 50.3\ \Omega$, the CPW's end is impedance matched. In this condition, the photon-magnon coupling strength decreases to $2.28$ MHz, and the frequency variation range of the PMSM correspondingly shrinks (Fig. \ref{fig2}(c)). As we further increase the VR to $R = 551\ \Omega$ with $\Gamma=0.83$, the photon-magnon coupling strength increases to $3.48$ MHz, broadening the frequency variation range of the PMSM (Fig. \ref{fig2}(d)). These results clearly demonstrate the in situ tunability of the effective coupling strength via the reflection coefficient.

To further quantify this evolution, we perform similar measurements at different values of $\Gamma$. We then extract the PMSM peak frequencies from these measurements and summarize them in Fig. \ref{fig3}(a). These data are fitted using Eq. \ref{equation3} and show excellent agreement with our model. For each $\Gamma$, the PMSM detuning $\Delta \omega_{s} / 2\pi$ exhibits a non-monotonic variation with the coil current, while the extent of this variation strongly depends on $\Gamma$, resulting in a pronounced U-shaped dependence of the tuning range on $\Gamma$. For clarity, the extreme points (local maxima) of each dispersion curve are selected (red stars connected by a dashed line) and projected onto the $YOZ$ plane. This projected variation directly reflects the evolution of the coupling strength, which decreases from $R = 551\,\Omega$ ($\Gamma \approx 0.83$) to $R = 50.3\,\Omega$ ($\Gamma \approx 0$) and then  rises again at lower resistance, reaching $R = 1.2\,\Omega$ ($\Gamma \approx -0.95$). 

Figure \ref{fig3}(b) shows the coupling strength $|g|/2\pi$ as a function of $\Gamma$. The dependence is non-monotonic, with a minimum near the impedance-matching condition (i.e., $\Gamma \approx 0$). The values increase toward both $\Gamma \rightarrow \pm 1$, which is consistent with the U-shaped behavior observed in Fig. \ref{fig3}(a). Fig. \ref{fig3}(c) shows the maximum slope of the PMSM frequency, $d(\omega/2\pi)/dI$, as a function of $\Gamma$. It exhibits a $\Gamma$-dependence similar to that of the coupling strength, demonstrating the tunability of our magnetometer via the VR.

The frequency evolution of the PMSM in our device is governed by the cubic equation (Eq. \ref{equation3}), which admits multiple solutions and therefore can exhibit either bistability or monostability depending on the polynomial coefficients (Eq. \ref{equation4}). Its bifurcation structure is determined by the coupling strength. The critical condition for the onset of bistability and the CP can be expressed in terms of a threshold coupling strength $|g|_{\mathrm{th}}$, which is derived from Eq.~\ref{equation3} as follows:
\begin{equation}
|g|^2_{\mathrm{th}} = \alpha_t^2 \csc^3\left( \frac{\pi}{6} + \frac{2}{3}\varphi' \right),
\label{equation5}
\end{equation}

\begin{figure}[htbp]
\includegraphics[width = 8.17 cm, trim=0.63cm 0.4cm 0.8cm 1.45cm, clip]{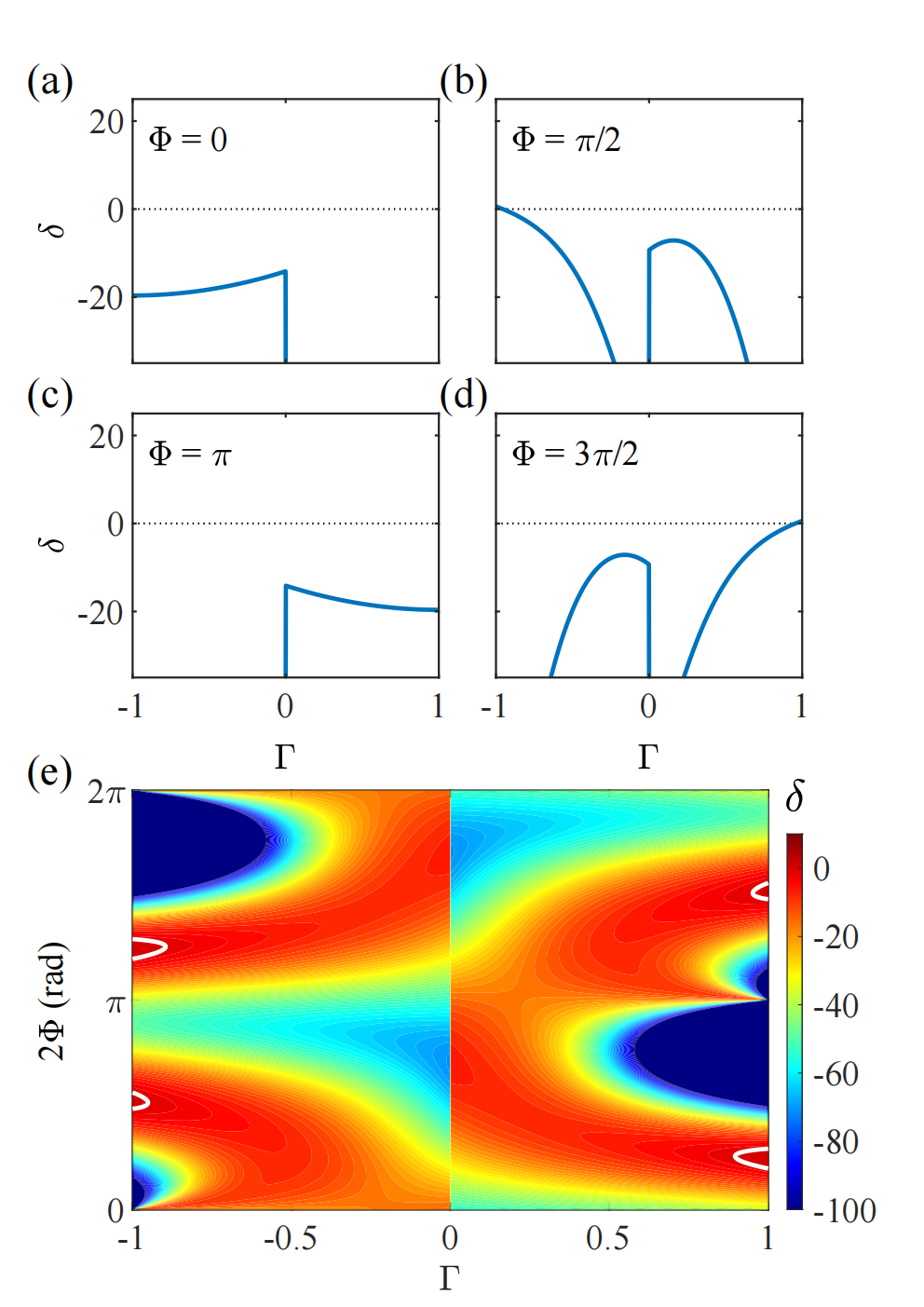}
\caption{
(a)-(d) Calculated effective coupling strength $|g|^2$ (solid blue lines) and critical condition $|g|^2_{th}$ (dashed orange lines) versus the reflection amplitude $\Gamma$ at four typical standing-wave phases $\varPhi = 0, \pi/2, \pi$, and $3\pi/2$, respectively. (e) Two-dimensional colormap mapping the net coupling parameter $\delta$ ($\delta = |g|^2 - |g|^2_{th}$) across the full $(\Gamma, \varPhi)$ parameter space. The dark red zones ($\delta > 0$) and the orange to blue zones ($\delta < 0$) denote the bistable and monostable regions, respectively, separated by the exact critical condition ($\delta = 0$).
}
\label{fig4}
\end{figure}

As the reflection amplitude $\Gamma$ is tuned by the VR, both the effective coupling strength $|g|$ and the threshold $|g|_{\mathrm{th}}$ vary simultaneously. Using Eq.~\ref{equation4} and Eq.~\ref{equation5}, their variations with $\Gamma$ are evaluated and plotted in Fig.~\ref{fig4}. Specifically, Figures \ref{fig4}(a)–(d) present the difference of $|g|^2$ and $|g|^2_{\mathrm{th}}$, i.e., $\delta=|g|^2-|g|^2_{\mathrm{th}}$, as functions of $\Gamma$ at four standing-wave phases. At $\varPhi = 0$ [Fig.~\ref{fig4}(a)] and $\varPhi = \pi$ [Fig.~\ref{fig4}(c)], the threshold diverges, yielding $\delta< 0$ over the entire range of $\Gamma$, and the system remains monostable. In contrast, near $\varPhi = \pi/2$ with $\Gamma = -1$ [Fig.~\ref{fig4}(b)] and $\varPhi = 3\pi/2$ with $\Gamma = 1$ [Fig.~\ref{fig4}(d)], the condition $|g|^2 \ge |g|^2_{\mathrm{th}}$ is satisfied, leading to bistable switching. For a global view, Fig.~\ref{fig4}(e) maps $\delta$ over the full $(\Gamma, \varPhi)$ parameter space. Regions with $\delta > 0$ and $\delta < 0$ correspond to bistable and monostable regimes, respectively, separated by the boundary $\delta = 0$ (white solid lines). This demonstrates continuous control of the system state via terminal impedance.

In conclusion, we construct a miniaturized PMSM magnetometer integrated on a PCB. By integrating a VR in the device, we enable in situ tuning of the photon–magnon coupling strength by controlling the reflection coefficient of microwave in the CPW. The coupling strength $|g|/2\pi$ can be tuned from $3.48$ MHz to a minimum value of $2.28$ MHz and subsequently increased to $3.27$ MHz by varying the terminal resistance from $551~\Omega$ to $1.2~\Omega$ (reflection coefficient from 0.83 to 0.95). Furthermore, we develop a theoretical model that incorporates the boundary-induced interference effect and allows us to derive the critical condition for the CP singularity. The theoretical predictions agree well with the experimental results, confirming the proposed coupling modulation mechanism. This work provides an effective approach for optimizing compact PMSM magnetometers and facilitates the development of portable and adaptive magnetic sensing platforms.

\begin{acknowledgments}
This work is financially supported by the Key R\&D Program of Shandong Province (Grant No. 2025CXPT204), the National Key Research and Development Program of China (Grant No. 2023YFA1406604), the STCSM (No. 23JC1404100), the National Natural Science Foundation of China (Grant Nos. 12522405, 12474120, 12504141), the Taishan Young Scholar Program of Shandong Province (Grant No. tsqn202507065), the Shandong Provincial Natural Science Foundation, China (Grant Nos. ZR2024YQ001 and ZR2025QC1475), and the Shandong Basic Physics Research Program (Grant No.TQ012025004).
\end{acknowledgments}

\section*{Data Availability}
The data that support the findings of this article are not publicly available. The data are available from the authors upon reasonable request.

\section*{Appendix: Equations of motion and derivation of the cubic equation}

The effective non-Hermitian Hamiltonian of the coupled photon–magnon system is given by
\begin{equation}
\mathcal{H} = \mathcal{H}_0/\hbar-i\Lambda,
\end{equation}
where $\mathcal{H}_{\mathrm{sys}}$ describes the bare photon and magnon modes, the gain term represents the saturable amplification in the photonic mode, and $\Lambda$ accounts for the radiation-induced coupling and dissipation modified by the reflection coefficient $\Gamma$.

Following the Heisenberg equations of motion,
\begin{equation}
i\hbar \frac{d\hat{a}}{dt} = [\hat{a},\mathcal{H}], 
\qquad
i\hbar \frac{d\hat{m}}{dt} = [\hat{m},\mathcal{H}],
\end{equation}
the dynamical equations can be explicitly written as
\begin{equation}
\begin{aligned}
\frac{d\hat{a}}{dt} 
&= -i(\omega_p - i\beta) \hat{a}
+ N\!\left(1 - i\sqrt{\frac{\epsilon}{2}}\hat{a} - \epsilon |\hat{a}|^2 \right)\hat{a} \\
&\quad - \kappa(1+\Gamma e^{i2\varPhi})\hat{a} - i\sqrt{\kappa\gamma}\,e^{i\varphi}(1+\Gamma e^{i2\varPhi})\hat{m},
\\[6pt]
\frac{d\hat{m}}{dt} 
&= -i(\omega_m-i\alpha) \hat{m}- \gamma(1-\Gamma e^{i2(\varPhi+\varphi)})\hat{m} \\
&\quad - i\sqrt{\kappa\gamma}\,e^{i\varphi}(1+\Gamma e^{i2\varPhi})\hat{a}.
\end{aligned}
\end{equation}

Under the inclusion of saturable gain, the system can reach a synchronized steady state. Neglecting the dynamical correction to the gain-induced damping, both modes oscillate at a common frequency $\omega_s$ with a fixed phase difference $\theta$. Based on this consumption, we can derive Eq. (\ref{equation3}). 

To further identify the saddle-node bifurcation (SNB) condition\cite{Strogatz2018,kuznetsov2023elements}, we examine the sensitivity of the synchronized frequency $\omega_s$ with respect to the control parameter $\omega_m$. Differentiating the cubic equation implicitly with respect to $\omega_m$ yields
\begin{equation}
\frac{d\omega_s}{d\omega_m}
= - \frac{\partial F / \partial \omega_m}{\partial F / \partial \omega_s},
\end{equation}
where $F(\omega_s,\omega_m)=0$ denotes the cubic state equation. The denominator explicitly reads
\begin{equation}
\frac{\partial F}{\partial \omega_s}
= 3\Delta\omega_s^2 + 2a_2 \Delta\omega_s + a_1.
\end{equation}

The SNB occurs when the slope diverges, i.e., when the denominator vanishes:
\begin{equation}
3\Delta\omega_s^2 + 2a_2 \Delta\omega_s + a_1 = 0.
\end{equation}

This condition defines the SNB boundaries in parameter space. Furthermore, the CP corresponds to the merging of the two SNB branches\cite{Strogatz2018}, which requires the simultaneous satisfaction of
\begin{equation}
\begin{cases}
F(\omega_s,\omega_m) = 0, \\
\partial F/\partial \omega_s = 0, \\
\partial^2 F/\partial \omega_s^2 = 0.
\end{cases}
\end{equation}

The second derivative is given by
\begin{equation}
\frac{\partial^2 F}{\partial \omega_s^2}
= 6\Delta\omega_s + 2a_2.
\end{equation}

Solving the above conditions yields
\begin{equation}
\Delta\omega_s = -\frac{a_2}{3},
\end{equation}
and substituting back leads to the critical condition for the CP. After parameter substitution, the threshold coupling strength is obtained as
\begin{equation}
|g|^2_{th} = \alpha_t^2 \csc^3\left( \frac{\pi}{6} + \frac{2}{3}\varphi' \right),
\end{equation}
which determines the onset of bistability and cusp singularity in the system.

\bibliographystyle{apsrev4-2}
\bibliography{ref1}

\end{document}